\documentclass[twocolumn]{aastex631}
\usepackage{multirow}
\usepackage{threeparttable}

\begin{document}

\title{A Comprehensive Catalog of Emission-line Nebulae, Star Clusters, and Supergiants in M31 from the LAMOST Spectroscopic Survey}

\author[0009-0007-5623-2475]{Pinjian Chen}
\affiliation{Department of Astronomy, Yunnan University, Kunming 650500, Yunnan, P.~R.\ China}
\affiliation{School of Astronomy and Space Science, University of Chinese Academy of Sciences, Beijing 100049, P.~R.\ China}
\affiliation{CAS Key Laboratory of Optical Astronomy, National Astronomical Observatories, Chinese Academy of Sciences, 20A Datun Road, Beijing 100101, P.~R.\ China}

\author[0000-0003-2472-4903]{Bingqiu Chen} 
\affiliation{South-Western Institute for Astronomy Research, Yunnan University, Kunming 650500, Yunnan, P.~R.\ China} 

\author[0000-0003-1286-2743]{Xuan Fang}
\affiliation{CAS Key Laboratory of Optical Astronomy, National Astronomical Observatories, Chinese Academy of Sciences, 20A Datun Road, Beijing 100101, P.~R.\ China}
\affiliation{School of Astronomy and Space Science, University of Chinese Academy of Sciences, Beijing 100049, P.~R.\ China}
\affiliation{Xinjiang Astronomical Observatory, Chinese Academy of Sciences, 150 Science 1-Street, Urumqi, Xinjiang 830011, P.~R.\ China}
\affiliation{Laboratory for Space Research, Faculty of Science, The University of Hong Kong, Pokfulam Road, Hong Kong, P.~R.\ China}

\author[0000-0003-2471-2363]{Haibo Yuan}
\affiliation{School of Physics and Astronomy, Beijing Normal University, Beijing 100875, P.~R.\ China}
\affiliation{Institute for Frontiers in Astronomy and Astrophysics, Beijing Normal University, Beijing 102206, P.~R.\ China}

\author[0009-0003-2888-6317]{Baisong Zhang}
\affiliation{South-Western Institute for Astronomy Research, Yunnan University, Kunming 650500, Yunnan, P.~R.\ China}

\author[0009-0000-7432-1390]{Xiangwei Zhang}
\affiliation{CAS Key Laboratory of Optical Astronomy, National Astronomical Observatories, Chinese Academy of Sciences, 20A Datun Road, Beijing 100101, P.~R.\ China}
\affiliation{School of Astronomy and Space Science, University of Chinese Academy of Sciences, Beijing 100049, P.~R.\ China}

\author[0000-0001-7169-4642]{Jiarui Sun}
\affiliation{Department of Astronomy, School of Physics and Astronomy, Shanghai Jiao Tong University, Shanghai 200240, P.~R.\ China}

\author[0000-0003-1295-2909]{Xiaowei Liu}
\affiliation{South-Western Institute for Astronomy Research, Yunnan University, Kunming 650500, Yunnan, P.~R.\ China}

\correspondingauthor{Bingqiu Chen, Xuan Fang, Xiaowei Liu}
\email{bchen@ynu.edu.cn, fangx@nao.cas.cn, x.liu@ynu.edu.cn}

\begin{abstract}
Spectroscopic observations of various tracers in nearby galaxies, such as Andromeda (M31), play a crucial role in identifying and classifying individual stellar populations and nebular objects, thereby enhancing our understanding of galactic composition, environment, and dynamics as well as stellar evolution.  While the LAMOST (Large Sky Area Multi-Object Fibre Spectroscopic Telescope) survey of M31 has produced extensive datasets, a comprehensive catalog of emission-line nebulae, star clusters, and supergiants is yet to be completed.  In this paper, we present a final catalog of 384 emission-line nebulae, 380 star clusters, and 375 supergiants and candidates in M31, as carefully selected and identified from the LAMOST spectroscopic database.  These objects were classified using a random forest algorithm, followed by thorough visual examinations of their spectral characteristics as well as morphologies revealed by archive images.  For emission-line nebulae, we measured radial velocities and relative fluxes of emission lines, enabling further classification of planetary nebulae and H~{\sc ii} regions.  Additionally, we identified 245 emission-line nebulae in M33.  This work lays the data foundation for the study of M31, and offers valuable tracers to investigate M31's structure and evolution. 

\end{abstract}


\section{Introduction} \label{sec:intro}

The Andromeda Galaxy (M31), our closest large disk neighbor, is located approximately 780\,kpc away \citep{Holland1998} and is the brightest and likely the most massive member of the Local Group.  At a slightly greater distance of 809\,kpc \citep{McConnachie2005}, the Triangulum Galaxy (M33) ranks as the third-largest member of the Local Group, after M31 and the Milky Way (MW). Their proximity and brightness make M31 and M33 ideal environments for studying astrophysical processes, including stellar evolution, the interstellar medium, and the formation and evolution of galaxies. Individual objects in these galaxies can be detected and studied with high precision, providing valuable insights into these processes. 

Identifying and classifying the various components of large spiral galaxies like M31 and M33 is essential for understanding galaxies similar to our own. While progress has been made over many decades, it remains a foundational challenge. Star clusters, for instance, have long been used as tracers of the global properties and formation history of M31. Edwin Hubble first identified 140 globular clusters (GCs) likely associated with M31 \citep{Hubble1932}. Later, \citet{Galleti2004} updated the Bologna catalog \citep{Battistini1987}, expanding it to include 693 known and candidate GCs, using data from the Two Micron All Sky Survey \citep[2MASS;][]{Cutri2003}, leading to the Revised Bologna Catalog (RBC). More recently, high-resolution images from the Pan-Andromeda Archaeological Survey \citep[PAndAS;][]{McConnachie2009} have further advanced the discovery of GCs and their candidates in M31 \citep{Huxor2014, Veljanoski2014}, with machine learning techniques offering even greater precision (\citealt{Wang2022}, \citealt{Wang2023}, Zhang in preparation).

Similarly, emission-line objects in M31 have been extensively cataloged. \citet[hereafter M06]{Merrett2006} compiled a comprehensive catalog of 3,300 emission-line objects in M31, of which 2,615 are likely planetary nebulae (PNe). A more recent imaging survey of a 54-square-degree area centered on M31 increased the total number of PNe and candidates to 5,625 \citep{Bhattacharya2019, Bhattacharya2021}, although only a fraction of these have been confirmed through spectroscopy \citep{Bhattacharya2022}. 

In addition, large photometric surveys, such as the Local Group Galaxies Survey \citep[LGGS;][]{Massey2006}, have provided opportunities to search for massive stars. Using radial velocity data, \citet{Massey2009} and \citet{Drout2009, Drout2012} successfully distinguished red and yellow supergiants in M31 and M33 from foreground MW stars. More recently, \citet[hereafter R21]{Ren2021} identified a significant sample of red supergiants in M31 and M33 using near-infrared color diagrams.

The development of wide-field, large-scale spectroscopic surveys has greatly expanded our ability to study extragalactic systems. The Large Sky Area Multi-Object Fiber Spectroscopic Telescope \citep[LAMOST;][]{Cui2012}, with its 4000 robotic fibers and 20\,deg$^2$ field of view, is particularly well-suited for spectroscopic surveys of individual objects in M31. Several pioneering studies have used LAMOST data to study emission-line nebulae \citep[e.g., PNe and H~{\sc ii} regions,][]{Yuan2010, Xiang2017, Zhang2020}, GCs \citep{Chen2015, Chen2016, Wang2021}, and massive stars \citep{Huang2019, Liu2022, Wu2024} in M31. However, although LAMOST has completed its first- and second-phase surveys, and M31 is not included in future observation plans, a comprehensive catalog of M31 objects from LAMOST data remains unpublished. This paper aims to address this gap.

Due to LAMOST's limiting magnitude, only the brightest objects in M31 can be observed spectroscopically within a reasonable time frame. Consequently, our study focuses on three types of objects: emission-line nebulae, star clusters, and supergiants. These objects are excellent tracers for studying the chemical composition, kinematics, and stellar populations of their host galaxies. We applied a Random Forest classifier for an initial categorization of these objects. For those that could not be definitively classified based on spectra alone, we conducted a kinematic analysis. During this process, we also identified several emission-line nebulae in M33 as part of the byproduct of this work.

\section{LAMOST Spectroscopic Survey Data} 
\label{sec:lamost_spectra}

The LAMOST pilot survey was completed in June 2012, followed by a five-year first-phase survey and a subsequent five-year second-phase survey.  In the first phase, only low-resolution ($R\sim$1800) spectroscopy was carried out, whereas in the second phase both low-resolution and medium-resolution ($R\sim$7500) observations were made.  Our study is based on the LAMOST DR9 v1.0 data release\footnote{\url{http://www.lamost.org/dr9/v1.0/}}, publicly available in 2022 April.  This data release includes both low- and medium-resolution spectra observed between 2011 October and 2021 June. 

For this work, we focuse on the low-resolution spectroscopic catalog, selecting the sky region within 15~degrees centred on M31.  This selection provided us with 648,485 spectra corresponding to 466,680 unique sources.  We also incorporated spectra from earlier LAMOST low-resolution catalogs. These include low signal-to-noise (S/N) spectra that the LAMOST 1D pipeline could not classify \citep[]{Luo2015}, leading them to be labeled as ``Unknown".  In some cases, nebular spectra with minimal continuum may have been misclassified by the pipeline, prompting us to search for potential emission-line nebulae within this subset. This additional step contributed 2,950 spectra from 2,721 unique objects in the vicinity of M31 and M33. 

In total, our initial sample consisted of 651,435 spectra from 469,401 distinct objects. These spectra cover a wavelength range of $\sim$3700-9100\,{\AA}, with a resolution of $R\sim$1800 at 5500\,{\AA}.

\begin{figure*}
\centering
\includegraphics[width=1.65\columnwidth]{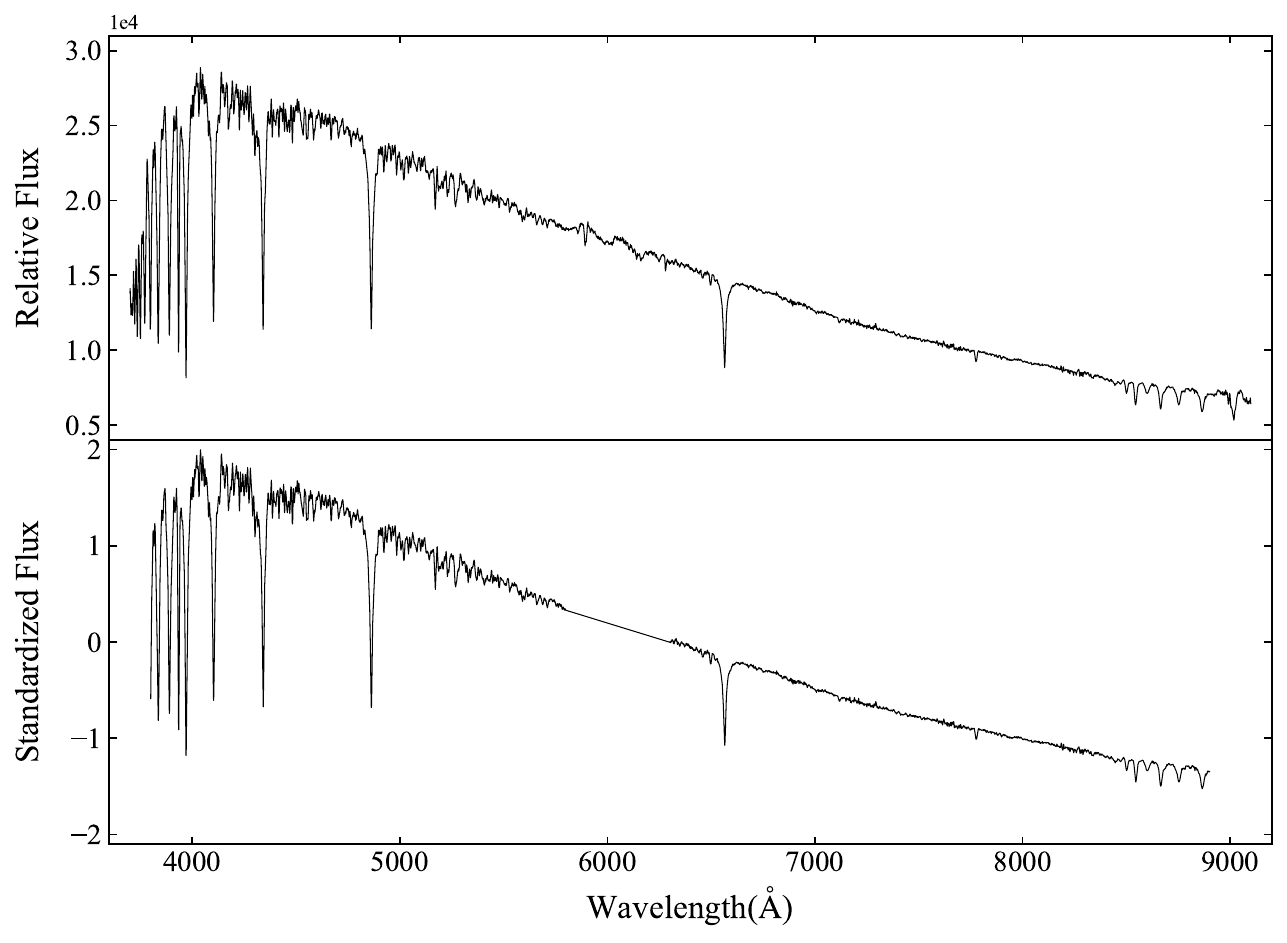}
\caption{Top panel: the original LAMOST spectrum of an A-type star.  Bottom panel: the same spectrum after preprocessing.  The spectrum in the wavelength region 5800--6300\,{\AA}, where the LAMOST blue and red arms are connected, is masked using a straight line. 
\label{fig:spectra}} 
\end{figure*}

\section{Initial Search for M31 Objects Using Random Forest Classifiers} \label{subsec:RF_classification}
The Random Forest (RF) algorithm is a robust supervised machine learning technique commonly used for both classification and regression tasks \citep{Breiman2001}. To identify potential candidates of M31 objects, we employed RF as the first step of our method, specifically utilizing the Random Forest Classifier (RFC). RF enhances predictive accuracy by utilizing a large ensemble of decision trees. Each tree is constructed using a randomly selected subset of the training data, generated through bootstrapping, and a randomly chosen subset of features. The model's performance is optimized by minimizing the error function (in this case, the Gini index), and the final prediction is made by averaging the results from all trees in the ensemble.

\subsection{The Training Sample} \label{subsec:training_sample}

The effectiveness of RFC depends on the quality of the training sample. To build a comprehensive and reliable dataset, we collected sources from several catalogs, ensuring both high completeness and minimal contamination. Our search extended beyond the SIMBAD database to include emission-line objects from M06, star clusters and candidates from the Revised Bologna Catalogue version 5 (RBC V5)\footnote{\url{http://www.bo.astro.it/M31/}} and \citet[hereafter C15]{Chen2015} , as well as red supergiants from R21. We also incorporated online resources such as ``PNe in M31'', ``HII regions in M31'', ``Just star clusters in M31'', and ``Stars in the M31 catalog'', all contributed by Nelson Caldwell\footnote{\url{https://lweb.cfa.harvard.edu/oir/eg/m31clusters/M31_Hectospec.html}, accessed on 4th September 2022.}.

We cross-matched these catalogs with our selected LAMOST catalog within a $3\arcsec$ radius and visually inspected each spectrum. Based on spectral features, we classified the objects into three categories: stars without emission lines (class 0), emission-line objects (class 1), and star clusters (class 2). Class 0 included both foreground Galactic stars and supergiants without emission lines from M31 and M33, while class 1 comprised emission-line nebulae and emission-line supergiants within these galaxies. Class 2 consisted of both young and old star clusters in M31 and M33. For candidate objects from the literature, we applied the same classification criteria as those used for confirmed objects.

We excluded from the training sample the LAMOST spectra with low S/N, missing data, or any contaminations. Additionally, objects labeled as ``Unknown'' were discarded at this stage due to the absence of radial velocity information. After this filtering process, our initial training sample comprised 38,522 spectra for class 0, 370 spectra for class 1, and 604 spectra for class 2.

\subsection{Data Preprocessing and Model Training} \label{subsec:data_preprocessing_model_training}

To input data into the RFC, the data must be numerical. In the current work, we used the fluxes at individual wavelengths as input features. First, we applied linear interpolation to the redshift-corrected LAMOST spectra, limiting the rest-frame wavelength range to $\rm 3800\,\AA$ to $\rm 8900\,\AA$. To avoid potential misalignments, we excluded the wavelength region between $\rm 5800\,\AA$ and $\rm 6300\,\AA$, where the blue and red arms of the spectrum connect. Subsequently, we performed standard z-score normalization, defined as:
\begin{equation}
    Z = \frac{F - \mu}{\rm \sigma},
\end{equation}
where $F$ represents the flux, $\mu$ is the mean flux of the spectrum, and $\rm \sigma$ is the standard deviation. This process transformed the original spectra into a 4100$\times$1 array. Fig.~\ref{fig:spectra} shows an example of both the original and preprocessed spectra.

RF performs optimally when trained on balanced datasets \citep{Breiman2001}. To achieve this, we randomly split the dataset into two parts: 80\% for training and 20\% for testing. To enhance the training set, we applied noise augmentation by adding Gaussian noise to the flux values based on the ``IVAR'' (inverse variance) values provided by LAMOST. For data points without inverse variance information, we used the mean value of the entire spectrum. The final balanced training set contained 30,820 spectra for class 0, 31,096 spectra for class 1, and 30,528 spectra for class 2.

\begin{table}
\setlength{\tabcolsep}{12.pt}
\caption{\label{tab:model} Precision, Recall, and F1 Scores of Our Trained RFC for All Classes}
\begin{center}
    \begin{tabular}{cccc}
        \hline
        \hline
        Class & Precision & Recall & F1 score\\
        \hline
        Class 0 & 0.94 & 0.94 & 0.94\\
        Class 1 & 0.91 & 0.93 & 0.92\\
        Class 2 & 0.91 & 0.89 & 0.90\\
        \hline
    \end{tabular}
    \end{center}
\end{table}

\begin{figure}
    \includegraphics[width=1.0\columnwidth]{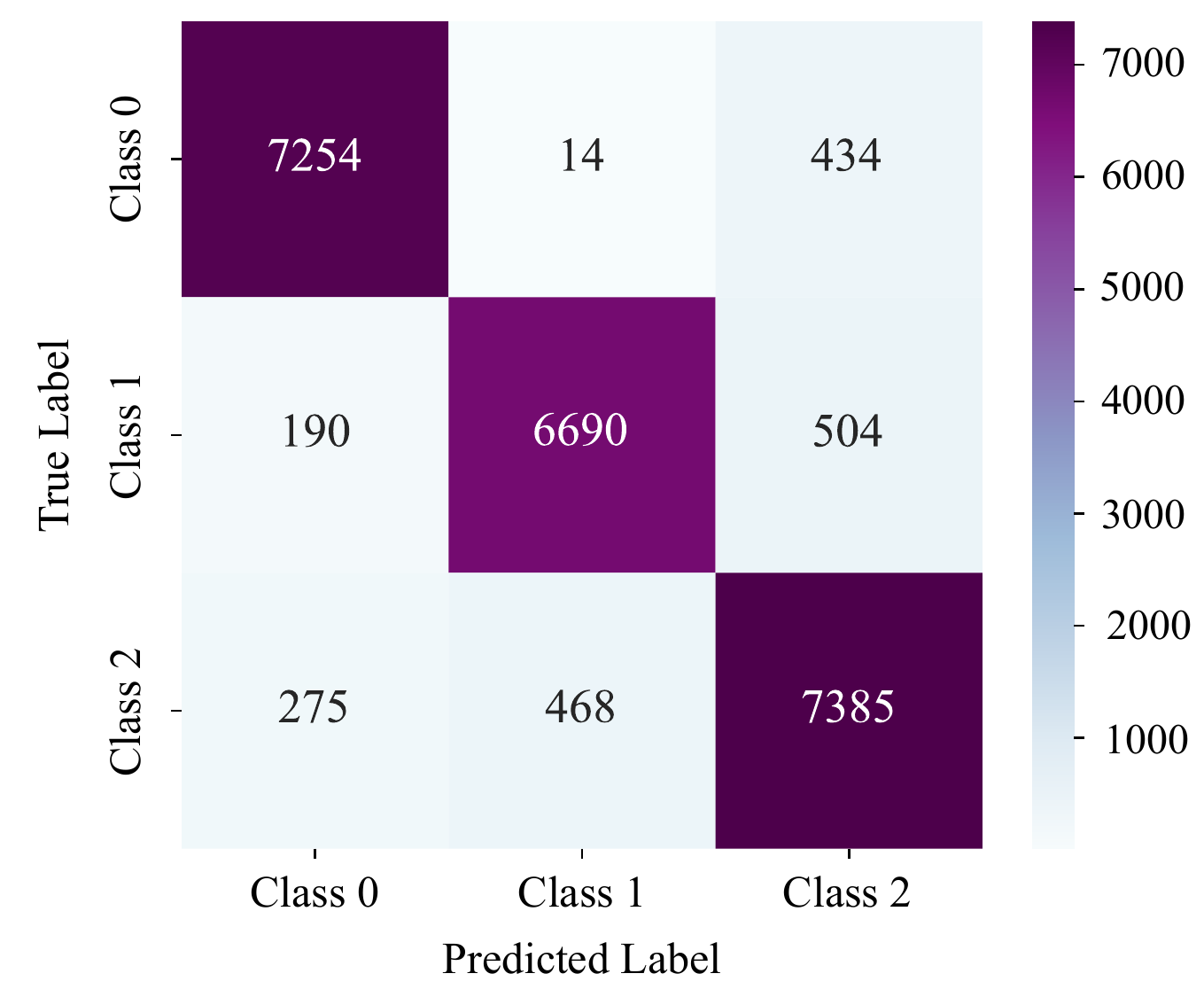}
     \caption{Confusion matrix showing the predicted and true classifications from the trained RFC.
    \label{fig:matrix}}
\end{figure}

We employ the \texttt{SCIKIT-LEARN} package for \texttt{PYTHON} \citep{scikit-learn} to train the RFC models. The \texttt{GridSearchCV} function is first adopted to optimize the most relevant hyperparameters for the Random Forest model. The final configuration included $ \emph{n\_estimators} = 500$, $\emph{max\_depth} = 30$, and $\emph{max\_features} = 500$. All other hyperparameters were left at their default values. To assess the model's performance, we used the F1 score and a confusion matrix. The F1 score is the harmonic mean of precision and recall, ranging from 0 to 1. For multi-class classification, the F1 score for each class is defined as:
\begin{equation}
     F1_i\ \text{score}= 2\cdot\frac{P_i \cdot R_i}{P_i + R_i},
\end{equation}
where $P_i$ is the precision and $R_i$ is the recall for class $i$. Precision and recall are given by:
\begin{equation}
        P_i = \frac{\mathrm{TP}_i}{\mathrm{TP}_i + \mathrm{FP}_i},\ \ \
        R_i = \frac{\mathrm{TP}_i}{\mathrm{TP}_i + \mathrm{FN}_i},
\end{equation}
where $\mathrm{TP}_i$ is the number of true positives for class $i$, $\mathrm{FP}_i$ is the number of false positives, and $\mathrm{FN}_i$ is the number of false negatives. The resulted precision, recall, and F1 scores of our trained RDC model for the individual classes are shown in Table~\ref{tab:model}, and the confusion matrix is displayed in Fig.~\ref{fig:matrix}. Our model achieved an overall accuracy of 92\%.

\begin{figure*}
\centering
    \includegraphics[width=1.75\columnwidth]{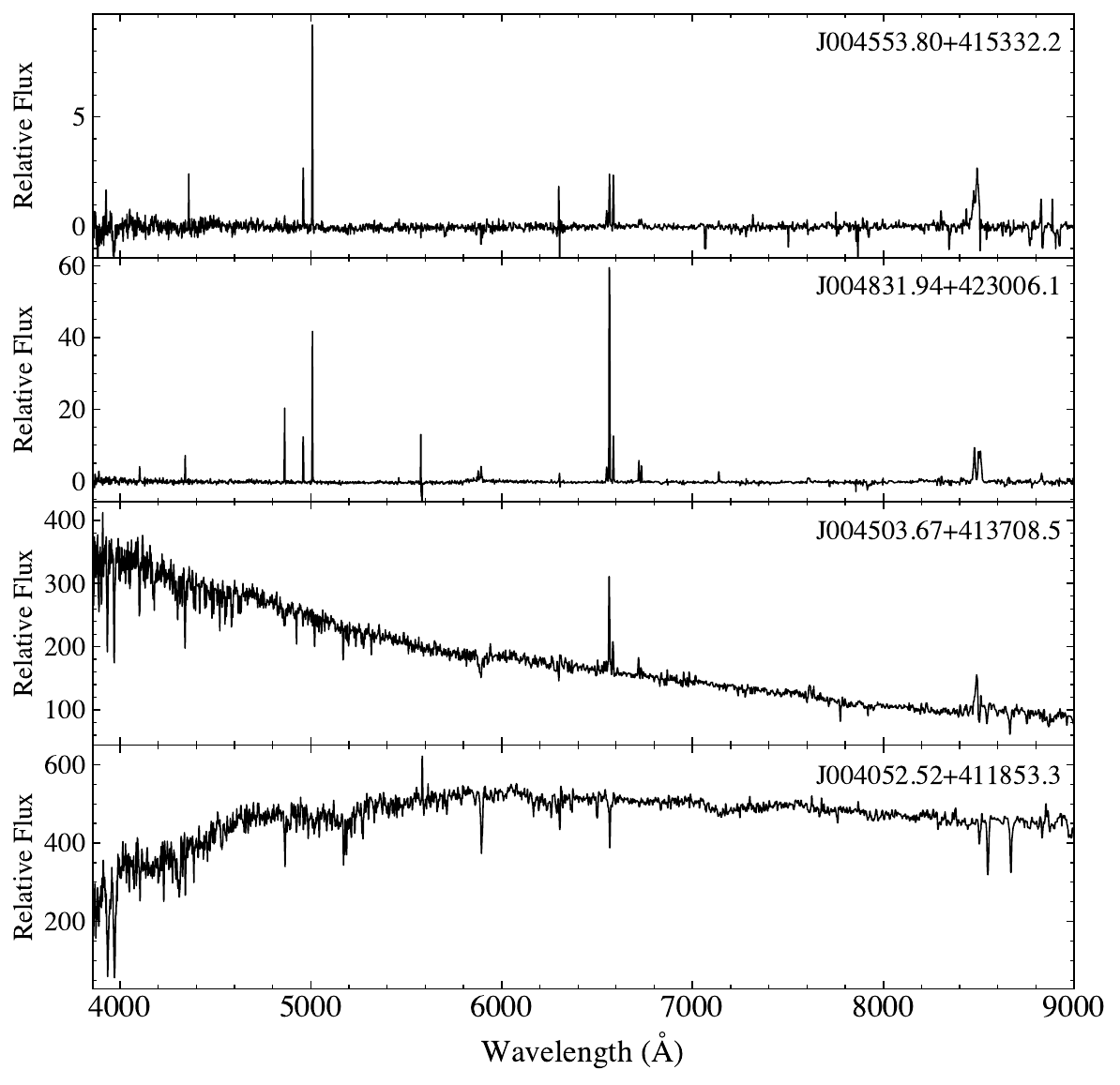}
     \caption{Examples of LAMOST spectra of the targets cataloged in this work, smoothed by 3 pixels.  From top to bottom: a PN candidate from M06, a newly discovered H~{\sc ii} region candidate, a newly discovered yellow supergiant candidate, and a known globular cluster from RBCV5. 
    \label{fig:spectra_example}}
\end{figure*}

\section{Further classification of candidates} 
\label{sec: searching for M31 members}

To minimize contamination from background galaxies and quasars, we excluded LAMOST spectra with radial velocities exceeding 2,000\,km\,s$^{-1}$. We also removed spectra labeled as ``Unknown" and those with erroneous redshift values (e.g., $-$9,999). These targets will be discussed in detail later (see Sect.~\ref{subsec:further_visual_examination}). The remaining spectra, which were not part of the training sample, were classified using our trained RFC model. This procedure resulted in the classification of 418,850 spectra as class 0, 9,284 as class 1, and 183,623 as class 2. Example spectra for each of these object types are shown in Fig.~\ref{fig:spectra_example}. 

We performed a preliminary review of the spectra classified as class 1. The majority of these spectra displayed prominent {H$\alpha$} emission lines, though some were affected by low S/N or cosmic rays. We first removed the low-quality spectra from further analysis. Additionally, we noted that a significant portion of the class 1 candidates were M-type dwarfs, characterized by strong TiO and VO molecular bands. These objects were identified as foreground stars, as their intrinsic faintness makes them too dim to be detected by LAMOST at the distance of M31. Consequently, we excluded these spectra from further consideration. 

Moreover, we identified a clustering of objects in the northern sky region of our sample. Upon reviewing available {H$\alpha$} images from the Virginia Tech Spectral Line Survey \citep[VTTS;][]{Dennison1998}, we concluded that the nebular lines in these spectra originated from the ionized interstellar medium (ISM) of the Galactic foreground. As a result, we discarded these targets with $\delta > 45^{\circ}$.

The remaining spectra were carefully inspected. In most cases, those exhibiting strong absorption lines were classified as emission-line stars, while spectra with minimal continuum were identified as emission-line nebulae. However, the relatively large fiber diameter of LAMOST introduces the possibility of contamination from the bright continuum background of M31, especially near the spiral arms and central bulge. Therefore, we allowed more leniency for objects with strong emission lines in these regions. Determining membership for emission-line stars based solely on spectra with relatively low S/N remains difficult, and further classification is postponed. This will be addressed alongside radial velocity data in Section~\ref{subsec:Kinematic_Analysis}. Ultimately, our analysis identified 128 unique objects as emission-line nebulae.

Unexpectedly, the RFC model classified a large number of objects as class 2, suggesting they may be star clusters. By applying a probability threshold of 0.95, we obtained a sample of 2,482 unique objects. For these candidates, we examined their morphologies using high-resolution $g$- and $i$-band images from the PAndAS archive. However, none of the sources exhibited obvious extended structures. Thus, the class 2 candidates are predominantly stars, particularly those without emission lines.

\subsection{Kinematic Selection of Supergiants in M31} 
\label{subsec:Kinematic_Analysis}

M31 provides an excellent environment for studying supergiant populations and testing massive star evolutionary models. The main challenge in identifying supergiants within M31 is distinguishing them from foreground MW stars. For red supergiants, the two-color method \citep[e.g.,][]{Ren2021, Massey2021} is effective in removing foreground stars. However, this approach is less effective for blue and yellow supergiants. For these objects, a common strategy is to first select candidates based on color-magnitude diagrams and then compare their radial velocities with expected values \citep[e.g.,][]{Massey2009, Drout2009, Massey2016}. In this study, we adopted this latter method to identify M31 supergiants.

The radial velocity of an object in the disk plane of a spiral galaxy can be estimated using the relation $V_{\rm r} = V_{\rm sys} + V(R_{\rm gal}) \sin\xi \cos\theta$ \citep{Ford1970}, where $V_{\rm sys}$ is the systemic radial velocity, $V(R_{\rm gal})$ is the circular velocity at a deprojected galactocentric distance $R_{\rm gal}$, $\xi$ is the angle between the line of sight and the normal direction of the galactic disk plane, and $\theta$ is the azimuthal angle in the disk plane.  With $\cos \theta = X/R_{\rm gal}$, where $X$ represents the position along the major axis, a flat rotation curve ($V$($R_{\rm gal}$) = const) results in a linear relationship between $V_{\rm r}$ and $X/R_{\rm gal}$.  To compute the expected radial velocity $V_{\rm exp}$, we used the relation from \cite{Massey2016}:
\begin{equation}\label{expected}
    V_{\mathrm{exp}} = -311.8 + 242.0(X/R_{\rm gal}),
\end{equation}
which was derived from fitting the radial velocities of a large sample of red supergiants in M31. This relation differs by roughly $17\,\mathrm{km~s^{-1}}$ from the result obtained by \cite{Drout2009}.

\begin{figure}
    \includegraphics[width=1.0\columnwidth]{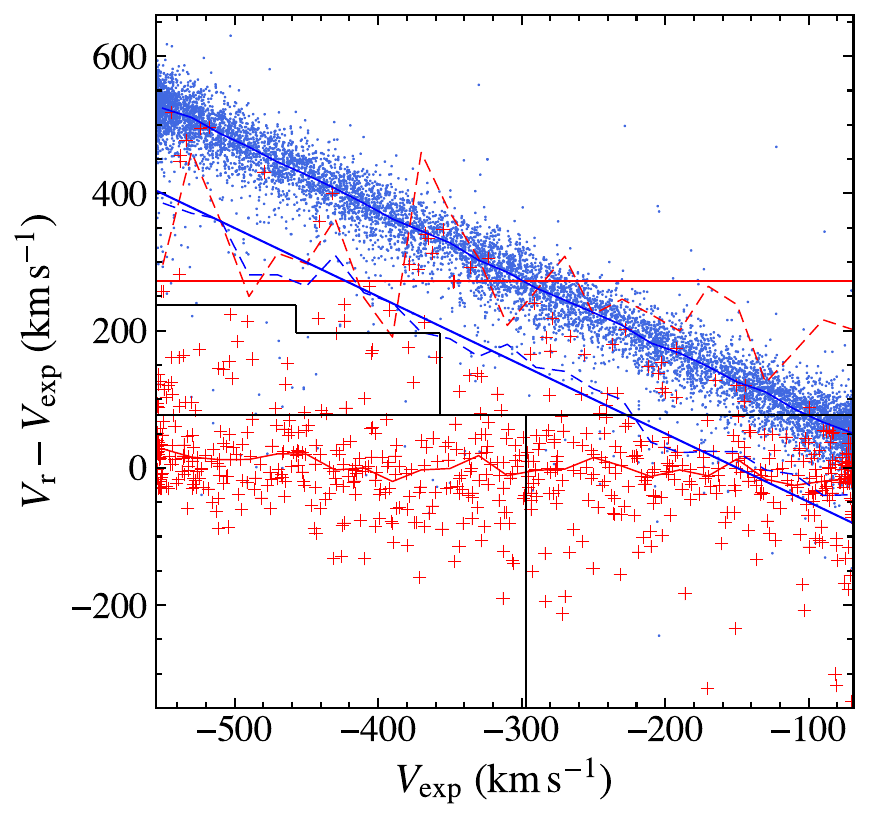}
    \caption{Comparison of the radial velocities, observed ($V_{\rm r}$) and expected ($V_{\rm exp}$), for M31 and MW objects, represented by red crosses and blue dots, respectively.  The velocity difference ($V_{\rm r}$ $-$ $V_{\rm exp}$) is plotted against $V_{\rm exp}$.  The median values of velocity difference for M31 and MW objects are shown by the red and blue solid curves, respectively, with the red- and blue-dashed curves representing the 3$\sigma$ scattering of data points.  The red-solid straight line indicates a constant $V_{\rm r} - V_{\rm exp}$ = 271.9\,km\,s$^{-1}$ for M31 objects, and the blue-solid straight line shows $V_{\rm r}$ = $-$149.5\,km\,s$^{-1}$ for the MW objects.  The black lines indicate the criteria (after correction) given by \citet{Drout2009}. 
    \label{fig:kinematics}}
\end{figure}

To assess the likelihood of an object belonging to M31 or the MW, we first compiled samples of both types from the LAMOST data. Known M31 objects were drawn from our catalogs of emission-line nebulae, star clusters, and stars, including supergiants and candidates identified in previous studies. These objects were required to have a galactocentric distance between 1 and 25\,kpc to ensure sufficient rotational support. We excluded ``high-velocity" tracers according to the criteria from \cite{Kafle2018}. For emission-line nebulae, we used the radial velocities measured in our analysis, while ``Unknown" nebulae were excluded due to their ambiguous nature. We visually inspected the radial velocities of all objects and removed any with poor-quality data. For objects with multiple observations, we selected the velocity measurement from the spectrum with the highest S/N. MW objects were selected from the SIMBAD database and classified as ``Star" with $R_{\rm gal}<$200\,kpc.  This yielded 661 M31 objects and 9,651 MW objects.

We calculated the expected radial velocities for all objects using Equation~\ref{expected}, and the parameters of M31 used in this calculation are listed in Table~\ref{tab:Parameters_of_M31}. Fig.~\ref{fig:kinematics} shows the kinematic distribution of these objects, along with the criteria from \cite{Drout2009}, adjusted by $17\,\mathrm{km~s^{-1}}$ for $V_{\rm exp}$. M31 objects, which are expected to exhibit significant disk rotation, should be located near the $y=0$ position, while foreground MW sources should align along the line representing $V_{\rm r} = 0$.

We divided the $V_{\mathrm{exp}}$ range from $-560$ to $-60\,\rm km~s^{-1}$ into 25 bins and calculated the median and standard deviation for both M31 and MW objects in each bin. These values were then used to define two Gaussian probability functions. The first Gaussian, ${\mathcal{N}_{\rm M31}}(\mu_1, \sigma_1^2)$, represents M31 objects with a mean $\mu_1 = 0.4\,\rm km~s^{-1}$ and a standard deviation $\sigma_1 = 90.5\,\rm km~s^{-1}$. The second Gaussian, ${\mathcal{N}_{\rm MW}}(\mu_2, \sigma_2^2)$, describes MW objects with a mean $\mu_2 = -V_{\rm{exp}} - 23.2\,\rm km~s^{-1}$ and a standard deviation $\sigma_2 = 42.1\,\rm km~s^{-1}$.

\begin{table}
\begin{center}
\setlength{\tabcolsep}{8.pt}
\caption{\label{tab:Parameters_of_M31} Parameters of M31 Adopted in This Work}
\begin{tabular}{lcc}
\hline
\hline
Parameter & Value & Ref. \\
\hline
R.A. (J2000) & $00^{\text{h}}42^{\text{m}}44.33^{\text{s}}$ & [1]\\
Decl. (J2000) & $+41^{\circ}16'07.5''$ & [1]\\
Position Angle & $38^{\circ}$ & [2]\\
Inclination Angle & $77^{\circ}$ & [3]\\
Distance &  780\,kpc & [4]\\
Heliocentric Radial Velocity & $-300\, \text{km}\,\mathrm{s^{-1}}$ & [5]\\
\hline
\end{tabular}
\end{center}
\tablenotetext{}{References: [1] \cite{Vaucouleurs1959}; [2] \cite{Kent1989}; [3] \cite{Walterbos1987}; [4] \cite{Holland1998}; [5] \cite{Vaucouleurs1991}}
\end{table}

Despite restricting our M31 sample to objects in the disk, some halo objects such as GCs and PNe located in the halo but projected onto the disk are inevitably included. These objects tend to exhibit a kinematic distribution characterized by a straight line, $V_{\rm r} - V_{\rm{exp}} = -V_{\rm{exp}} - 300\,\mathrm{km~s^{-1}}$, with large dispersion. This is evident in Fig.~\ref{fig:kinematics}, where red pluses appear in the upper-left and lower-right quadrants. A few M31 stars are also found in the upper-left region, likely belonging to the MW based on their kinematics (see Sect.~\ref{subsec: supergiant candidates}). Similarly, some MW objects may be misclassified as M31 stars in the SIMBAD database, but the uncertainty introduced by such misclassifications is expected to be less than $10\,\rm km~s^{-1}$ and is considered negligible.

Supergiants in M31, located near the top of the Hertzsprung-Russell (HR) diagram, are among the few bright stars detectable by LAMOST due to its magnitude limit ($r \sim 19\, \rm mag$). To identify supergiant candidates, we selected objects classified as class 1 and class 2 by the RFC model, along with those identified as emission-line stars. We also included objects classified as ``Star" in the SIMBAD database due to their ambiguous membership. However, we excluded 14 objects previously identified as foreground dwarfs or subgiants by \cite{Gordon2016}. All objects were required to have a galactocentric distance within 25\,kpc to the center of M31. The probabilities of each object belonging to M31 or the MW were estimated using the Gaussian probability density functions and scaled to a range of 0 to 1. Supergiant candidates were selected based on the following criteria:
\begin{enumerate}
    \item $P_{\rm M31}$ falls within the $3\sigma$ range of ${\mathcal{N}_{\rm M31}}(\mu_1 , \sigma_1 ^{2})$,
    \item $P_{\rm MW}$ lies outside the $3\sigma$ range of ${\mathcal{N}_{\rm MW}}(\mu_2 , \sigma_2 ^{2})$.
\end{enumerate}
Using these criteria, we identified 248 unique supergiant candidates. 

Though LAMOST also covers M33, the smaller systemic velocity of M33 \citep[$\sim -179\,\rm km\,s^{-1}$;][]{Vaucouleurs1991} reduces the ``blank" region between the red and blue curves in Fig.~\ref{fig:kinematics}, which we use to distinguish between M31 objects and MW stars. As a result, this method would likely suffer from significant foreground contamination, so we do not extend our supergiant selection to M33 in this work.

Furthermore, we note that the kinematic method used to select supergiant candidates is ineffective in a small region of M31's northeastern disc.  In this area, the expected radial velocities overlap with the velocities of foreground disk stars (see Fig.~\ref{fig:kinematics}). To address this, we visually examined the LAMOST spectra in this region. Fortunately, we identified a yellow supergiant candidate, LAMOST J004503.67$+$413708.5, based on its spectral features. The spectrum has sufficient S/N to clearly reveal luminosity-sensitive lines, including the \ion{Ti}{2} and \ion{Fe}{2} blends, as well as the \ion{O}{1} $\lambda$7774 triplet (see Fig.~\ref{fig:spectra_example}).

\subsection{Further Visual Examination of the ``Unknown'' Spectra Classified by LAMOST} 
\label{subsec:further_visual_examination}

The primary objective of this step is to identify the emission-line nebulae with their spectra classified as ``Unknown" by the LAMOST pipeline.  These spectra could not be categorized by the RFC model, as the absence of reliable redshift information prevents us from shifting them to the specific rest-frame wavelength range required for our analysis. Most of these spectra exhibit very low S/N and could not be successfully cross-matched with LAMOST spectral templates. Therefore, we focused only on spectra with strong emission lines.

Following a thorough visual inspection, we unexpectedly identified 309 emission-line nebulae in the vicinity of M31 and M33.  This numner closely matches the total number of the nebulae identified in the DR9 sample.  Moreover, we discovered seven objects projected onto the disc of M31 that exhibit H$\alpha$ emission lines with broad wings in their spectra.  This feature in line profile is commonly observed in massive supergiants and is attributed to Thomson scattering and other related mechanisms \citep{Bernat1978, Hillier1991}. Based on these characteristics, we propose these seven objects as supergiant candidates. 

To distinguish between previously known and newly identified objects, all candidates selected from the LAMOST ``Unknown" spectra were cross-matched with the catalogs described in Section \ref{subsec:training_sample}, following the same procedure.

\begin{splitdeluxetable*}{lccccccBcccccccc}
\tabletypesize{\scriptsize}
\setlength{\tabcolsep}{12.pt}
\tablecaption{Emission-line Nebulae in M31 and M33 Identified from the LAMOST Database \label{tab:nebula}}
\tablehead{
\colhead{Designation} & \colhead{R.A.} & \colhead{Decl.} & \colhead{M06 ID$^{a}$} & \colhead{M06\,$V_{\rm r}\,^{b}$} & \colhead{B24 ID$^{c}$} &\colhead{$V_{\rm r}$} & \colhead{blue $V_{\rm r}$} & \colhead{red $V_{\rm r}$} & \colhead{N2} & \colhead{O3} & \colhead{Host} & \colhead{Literature$^{d}$} & \colhead{Classification$^{e}$} & \colhead{Comments} \\
\colhead{} & \colhead{(J2000)} & \colhead{(J2000)} & \colhead{} & \colhead{(km\,s$^{-1}$)} & &\colhead{(km\,s$^{-1}$)} & \colhead{(km\,s$^{-1}$)} & \colhead{(km\,s$^{-1}$)} & \colhead{} & \colhead{} & \colhead{} & \colhead{} & \colhead{} & \colhead{}
}
\startdata
J004553.35+420850.4 & 11.47233 & 42.14734 & 32 & $-$70.3 & &$-72.1 \pm 9.6$ & $-72.0 \pm 10.0$ & $-74.2 \pm 34.7$ & $-$2.0 & 2.0 & M31 & PN & PN & ~ \\
J004634.35+421143.2 & 11.64315 & 42.19534 & 57 & $-$51.9 & 287 &$-74.1 \pm 0.2$ & $-63.0 \pm 0.3$ & $-95.5 \pm 0.4$ & $-$0.82 & 0.43 & M31 & HII & HII & ~ \\
J004610.59+421244.3 & 11.54413 & 42.21231 & 60 & $-$66.3 & &$-76.0 \pm 0.3$ & $-73.6 \pm 0.3$ & $-113.9 \pm 1.3$ & $-$0.7 & 0.2 & M31 & HII & HII & ~ \\
... & ... & ... & ... & ... & ... & ... & ... & ... & ... & ... & ... & ... & ... & ... \\
J004024.70+413727.9 & 10.10292 & 41.62444 & 3222 & $-$224.1& & $-222.0 \pm 1.7$ & $-223.1 \pm 1.9$ & $-217.0 \pm 4.0$ & $-$0.83 & 1.41 & M31 & PN & PN & NGC 205? \\
J004257.00+405101.8 & 10.73750 & 40.85050 & 3239 & $-$160.1& & $-155.2 \pm 2.2$ & $-151.5 \pm 3.7$ & $-157.2 \pm 2.8$ & $-$0.1 & 0.58 & M31 & PN\_c & PN\_c & M32? \\
J004047.70+413729.4 & 10.19875 & 41.62486 & 3247 & $-$387.1& & $-380.2 \pm 29.3$ & $-380.2 \pm 29.3$ & ~ & ~ & 2.0 & M31 & PN\_c & PN\_c & NGC 205? \\
... & ... & ... & ... & ... & ... & ... & ... & ... & ... & ... & ... & ... & ... & ... \\
J013255.98+303426.7 & 23.23326 & 30.57410 & ~ & ~ & &$-137.1  \pm 1.7$ & $-128.2 \pm 7.5$ & $-137.6 \pm 1.8$ & $-$0.54 & $-$2.0 & M33 & ~ & HII\_c & ~ \\
J013301.32+303044.1 & 23.25553 & 30.51226 & ~ & ~ & &$-128.3 \pm 0.4$ & $-127.1 \pm 0.7$ & $-128.7 \pm 0.4$ & $-$0.63 & 0.32 & M33 & HII & HII & ~ \\
J013415.68+303346.1 & 23.56536 & 30.56281 & ~ & ~ & &$-164.6 \pm 1.2$ & $-167.4 \pm 4.5$ & $-164.4 \pm 1.3$ & $-$0.58 & $-$0.18 & M33 & ~ & HII\_c & ~ \\
... & ... & ... & ... & ... & ... & ... & ... & ... & ... & ... & ... & ... & ... & ... \\
\enddata
\tablenotetext{a}{Object ID given in \citet{Merrett2006}.}
\tablenotetext{b}{Radial velocity given in \citet{Merrett2006}.}
\tablenotetext{c}{Object ID given in \citet{Bosomworth2024}.}
\tablenotetext{d}{Classification from the literature; candidate objects are marked as ``$\ast$\_c''.}
\tablenotetext{e}{New classification made in this work based on the LAMOST database.}
\tablecomments{This table is published in its entirety online only in the machine-readable format.}
\end{splitdeluxetable*}

\section{Results and Discussion} 
\label{sec:Results_and_Discussion}

\subsection{Emission-Line Nebulae} \
\label{subsec:Emission-Line_Nebulae}

We compiled a comprehensive catalog of emission-line nebulae from LAMOST data, encompassing 629 unique objects. In M31, the sample includes 102 known planetary nebulae (PNe), 122 PN candidates, 62 previously identified H~{\sc ii} regions, 76 candidates, and 22 unclassified nebulae. In M33, the catalog includes 30 known PNe, 15 candidates, 95 known H~{\sc ii} regions, 88 candidates, and 17 unclassified emission-line nebulae. Table~\ref{tab:nebula} lists each object's ID, position, radial velocity, emission-line ratio, and final classification.

Given the relatively low S/N of many LAMOST spectra, we conservatively labeled new discoveries as candidates, though some classifications are highly plausible. After cross-referencing with the SIMBAD database and relevant literature, we identified 32 PN candidates and 70 candidate H~{\sc ii} regions near M31 as new discoveries. In M33, we found 6 PN candidates and 84 H~{\sc ii} region candidates.  The term ``new" refers to objects that, to the best of our knowledge, have no prior identification within a 3\,$''$ radius. For some ambiguous objects from the literature, we updated their classifications based on LAMOST spectra. For instance, several PN candidates from M06 were reclassified as more likely being H~{\sc ii} regions. Recently, \citet{Alexeeva2022} presented a catalog of 95 H~{\sc ii} regions and 15 PNe in M33 using LAMOST DR7 data. While our catalog further expands the number of emission-line nebulae in M33, it includes the majority of objects in their work, with only a few excluded due to broad emission lines, suggesting they may be massive supergiants. 

Another more recent catalog was reported by \citet{Bosomworth2024}, who spectroscopically identified 294 H\,{\sc ii} regions in M31 using the Hectospec spectrograph on the 6.5-m Multi-Mirror Telescope (MMT).  We cross-matched our catalog with that of \citet{Bosomworth2024} and found 26 objects in common, of which 24 have consistent classifications.  Of the remaining two objects, one was classified as an ``Unknown'' emission-line nebula due to low S/Ns of its LAMOST spectrum, and the other one has been classified as a PN candidate based on its strong [N~{\sc ii}] nebular emission lines.  However, the classification of this object might be affected by contamination from the emission of diffuse ionized gas, as the object lies very close to the bulge region of M31.  These overlapping objects are marked in Table\,\ref{tab:nebula}.

\begin{figure}
 \centering
    \includegraphics[width=1.0\columnwidth]{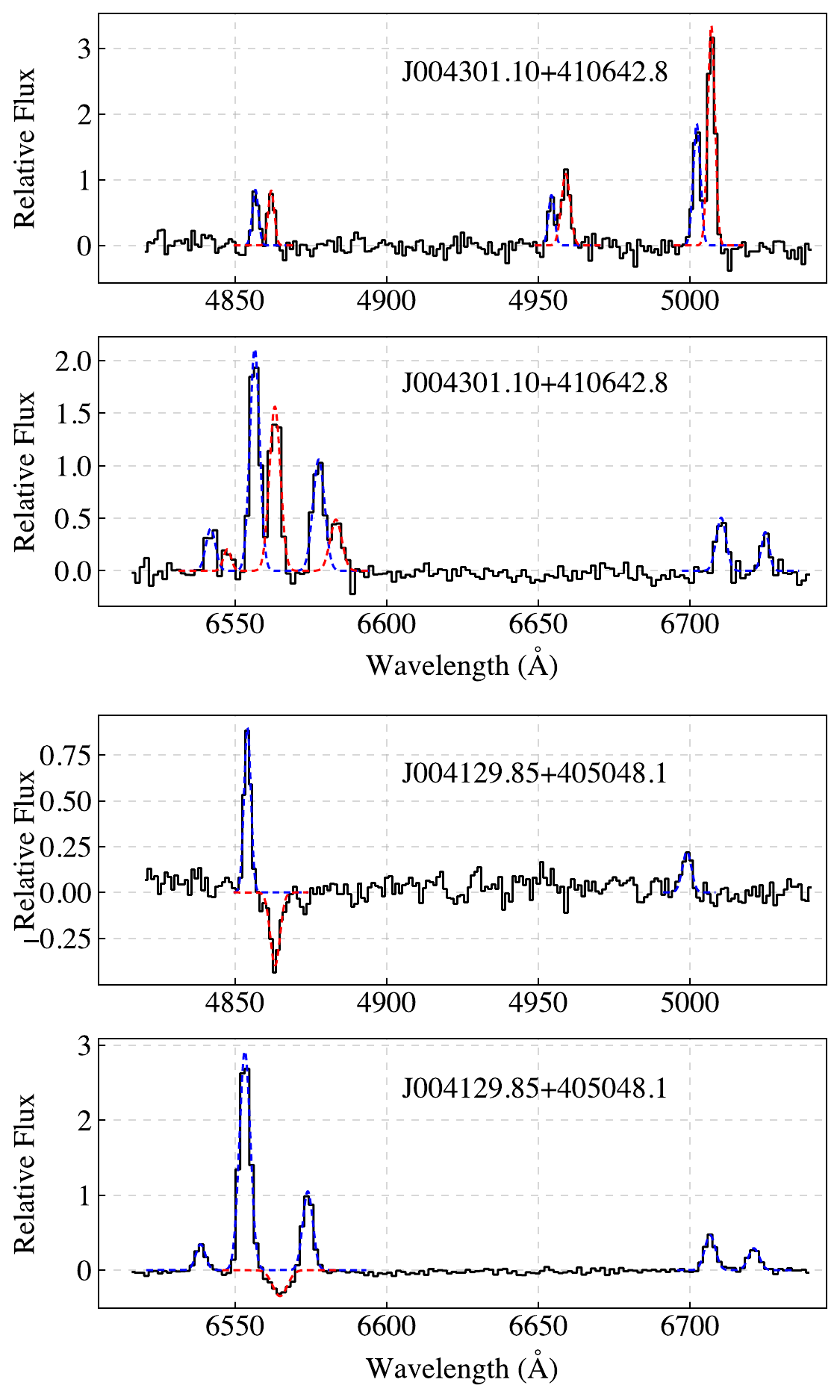}
     \caption{LAMOST spectra of two dual-velocity objects.  The two velocity components are color-coded for clarity: blue indicates the blueshifted component, while red signifies the redshifted.  Object J004301.10$+$410642.8 (top two panels), a PN candidate from M06, shows a velocity disparity of $\sim$280\,km\,s$^{-1}$.  Object J004192.85$+$405048.1 (bottom two panels), identified as an H~{\sc ii} region in M31, displays a velocity difference of 528\,km\,s$^{-1}$. 
     \label{fig:line-split}}
\end{figure}

\subsubsection{Relative Line Fluxes and Radial Velocities} \label{subsubsec:Line_Fluxes_and_Radial_Velocities}

To characterise spectral features and refine the classification of emission-line nebulae, we measured the radial velocities and fluxes of several emission lines for each object in our catalog. These lines include H$\beta$, [O~{\sc iii}] $\lambda\lambda$4959,\,5007, H$\alpha$, [N~{\sc ii}] $\lambda\lambda$6548,\,6584, and [S~{\sc ii}] $\lambda\lambda$6716,\,6731, when available.  A linear continuum was fitted within a $\sim$20\,{\AA} range, and each line was modeled with a Gaussian profile. The central wavelength of the fit determined the line's radial velocity, while line fluxes were extracted by integrating the area under the Gaussian after subtracting the continuum.

In regions where emission lines are blended (e.g., $\rm 6545 \sim 6590\,\AA$), we applied a method similar to \cite{Prichard2017}, fitting simultaneous Gaussian profiles using the equation:
\begin{eqnarray} \label{Gaussian}
   f(x) &= (&mx+c) + \frac{A_1}{\sigma_1\sqrt{2\pi}}\exp \left(-\frac{(x-\lambda_1)^2}{2\sigma_1^2}\right) \nonumber \\
   && +\frac{A_2}{\sigma_2\sqrt{2\pi}}\exp \left(-\frac{(x-(\lambda_1+\Delta \lambda))^2}{2\sigma_2^2}\right),
\end{eqnarray}
where $m$ and $c$ describe the continuum, $A_1$ and $A_2$ are the amplitudes of the two peaks, and $\sigma_1$ and $\sigma_2$ represent their widths. $\lambda_1$ is the central wavelength of the first emission line, and $\Delta \lambda$ is the separation between the two lines, based on the systemic velocities of M31 and M33. This method proved more accurate than fitting each line independently, as it reduced the number of free parameters, though slight velocity shifts might be introduced.

Given LAMOST's large fiber diameter, contamination from external sources, such as foreground stars or diffuse ionized gas in M31, is inevitable.  This issue is particularly pronouncing near the spiral arms and bulge regions, where contamination may render line-fitting unreliable. Occasionally, a secondary velocity component is evident in the spectra, as shown in Fig.~\ref{fig:line-split}. In such cases, we provided comments in the catalog and reported the velocities and fluxes consistent with previous literature.

The final radial velocity for each object was calculated as the weighted average of the velocities from individual emission lines, with weights inversely proportional to the error in each measurement. In some cases, discrepancies between the blue- and red-arm spectra radial velocities were observed, likely due to LAMOST's wavelength calibration. To account for this, we separately provided velocities derived from the blue- and red-arm spectra, ensuring careful interpretation of these values. A comparison with M06 velocities for 171 common objects revealed a mean difference of $-$3.54\,km\,s$^{-1}$ and a standard deviation of 18.28\,km\,s$^{-1}$.

\begin{figure}
\centering
    \includegraphics[width=1.0\columnwidth]{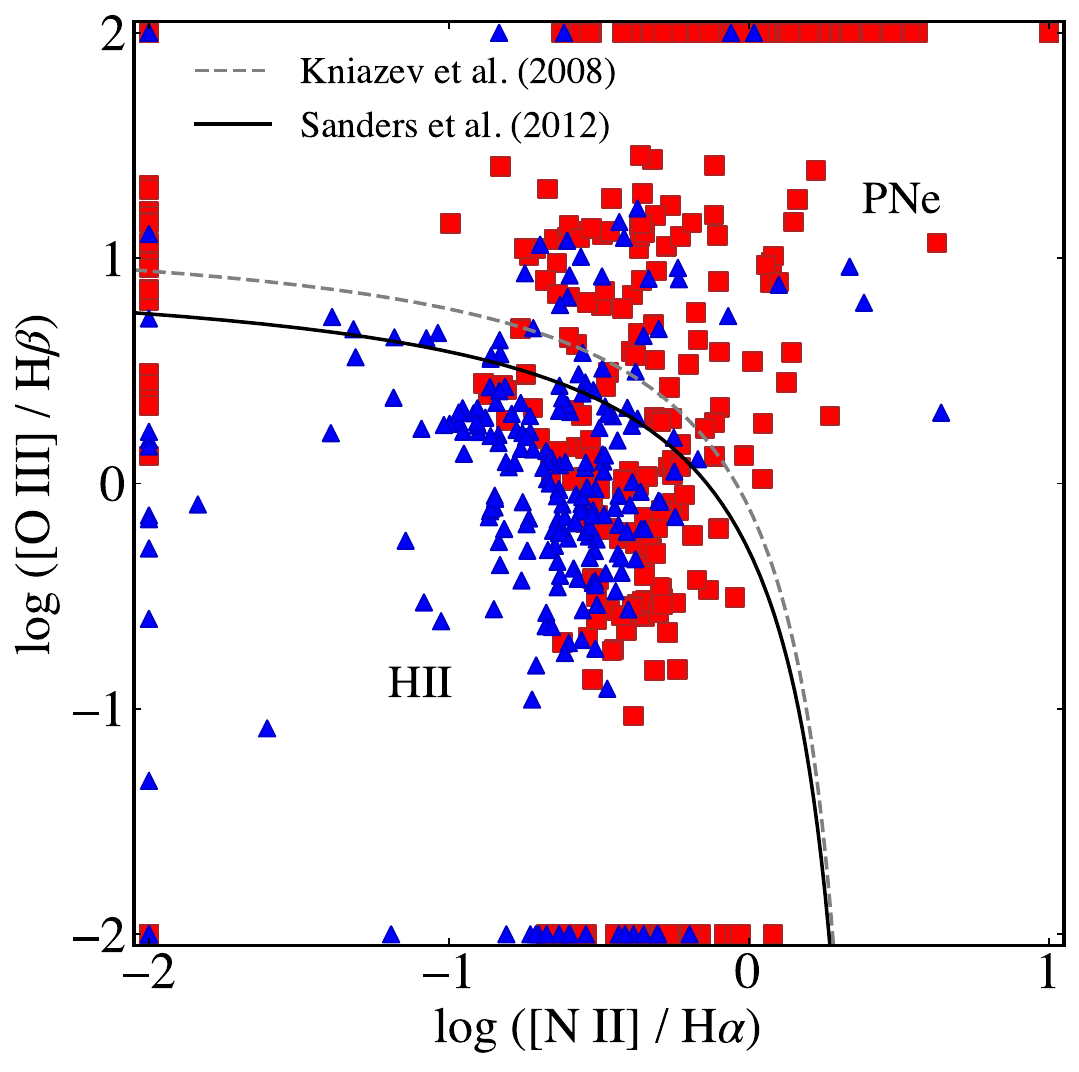}
     \caption{BPT diagram of the emission-line nebulae in our catalog.  The nebulae in M31 are shown as red squares, and those in M33 as blue triangles.  The black solid line represents our criterion for distinguishing PNe from H~{\sc ii} regions.  The grey dashed line reflects the demarcation given by \citet{Kniazev2008}.
    \label{fig:BPT}}
\end{figure}

\subsubsection{Final Classification} \label{subsubsec:Classification}

At a distance of approximately 800\,kpc, most emission-line nebulae appear as point sources in LAMOST data, except for a few extended H~{\sc ii} regions. This complicates morphological classification, making spectral features essential for identification. These features include forbidden emission lines from elements like oxygen, nitrogen, and sulfur, as well as hydrogen recombination lines. We used the Baldwin-Phillips-Terlevich \citep[BPT;][]{Baldwin1981} diagram (Fig.~\ref{fig:BPT}), which differentiates PNe from H~{\sc ii} regions based on the line ratios of [O~{\sc iii}] $\lambda 5007 / \rm{H}\beta$ (O3) and [N~{\sc ii}]$/$$\rm{H}\alpha$ (N2). Here, [N~{\sc ii}] represents the combined flux of the lines at $6548\, \rm{\AA}$ and $6583\, \rm{\AA}$.

While other diagnostic diagrams exist \citep{Frew2010}, incorporating red sulfur lines at $\lambda\lambda 6716,6731\,\rm{\AA}$, these lines are often weak in LAMOST spectra, which could introduce ambiguity due to low S/N. The BPT diagram is preferred because it uses lines of similar wavelengths, minimizing the effects of extinction, reddening, and flux calibration errors.  As shown in Fig.~\ref{fig:BPT}, some sources are plotted on the boundaries of the BPT diagram. This distribution is primarily due to the very low emission-line strengths observed in their LAMOST spectra. For example, sources found at the lower edge of the diagram exhibit [O~{\sc iii}] $\lambda 5007$ line intensities close to zero, resulting in O3 values near zero, which places them at the lower boundary. Conversely, sources at the upper edge of the diagram show almost no detectable H$\beta$ emission, causing their O3 values to approach infinity, and we have plotted them accordingly at the upper boundary of the diagram. 

To distinguish the PNe and  H~{\sc ii} regions from the BPT diagram, we adopted the criterion from \cite{Sanders2012}, specifically:
\begin{equation}\label{BPT_criteria}
    \rm O3 > (0.61/(N2-0.47))+1.0.
\end{equation}
In addition to using the BPT diagram, we visually inspected objects near the classification boundary to confirm their final types. As a result, 321 objects were classified as H~{\sc ii} regions and 269 as PNe. A small number of objects could not be classified due to low S/N, particularly in the blue-arm spectra, and were labeled as ``Unknown" in our catalog.

It's important to note that H~{\sc ii} regions often coincide with young, massive blue stars, which can raise the blue-arm continuum and introduce absorption components into the Balmer emission lines. This effect slightly reduces the measured intensities of these lines. Given LAMOST's $\sim 3''$ fiber diameter, this issue is exacerbated. However, we believe the classification remains reliable, as the Balmer emission lines in these spectra are generally much stronger than the [O~{\sc iii}] $\lambda\lambda 4959,5007$ lines, indicating low-excitation environments.

\begin{figure*}
\centering
    \includegraphics[width=2.1\columnwidth]{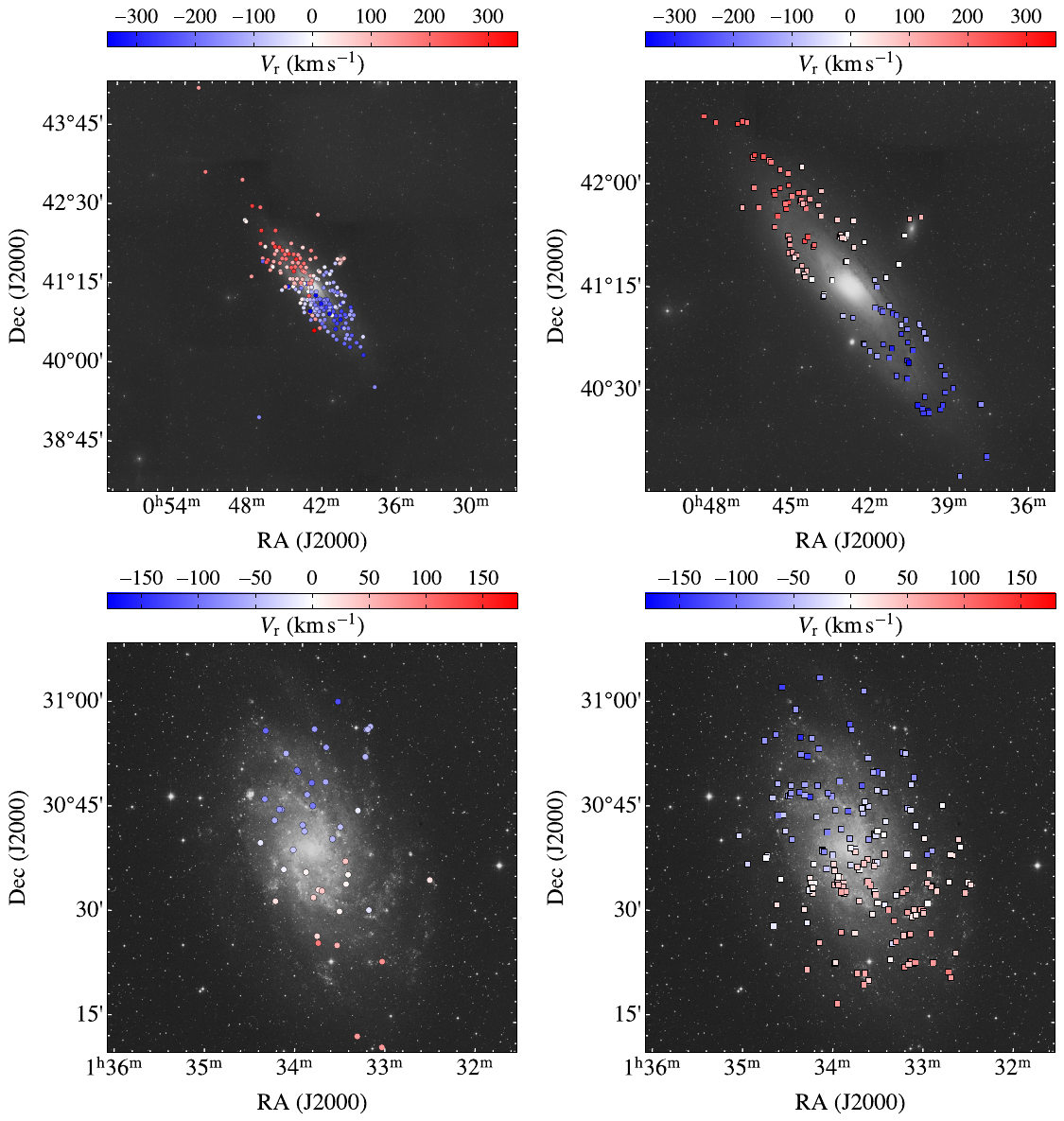}
     \caption{Spatial distributions of classified emission-line nebulae observed with LAMOST, overlaid on the mosaicked images of M31 and M33 as retrieved from the Digitized Sky Survey.  Top panels: spatial distributions of PNe (left panel, little circles) and H~{\sc ii} regions (right panel, little squares) in M31.  Bottom panels: same as top, but for M33.  Color-coding of the data symbols represents weighted radial velocity after subtracting the systemic velocity ($V_{\rm r}$) of the host galaxy.  A systemic velocity of $-$179 km\,s$^{-1}$ for M33 was adopted.
     \label{fig:spatial_distribution}}
\end{figure*}

\subsubsection{Spatial Distribution} 
\label{subsubsec:Spatial_Distribution}

Fig.~\ref{fig:spatial_distribution} shows the spatial distribution of all classified emission-line nebulae. In M31, PNe are spread across a wide region, from the bulge to the halo. Several PNe are spatially associated with known substructures revealed by the PAndAS survey \citep{McConnachie2018}, such as the Giant Stellar Stream and the Northern Clump. H~{\sc ii} regions trace the spiral structure of M31, particularly the star-forming ring at a radius of about $10\,$kpc. The velocity distribution of H~{\sc ii} regions closely follows M31's rotation, while PNe exhibit a more random distribution, reflecting a higher level of asymmetric drift.

We also identified a few objects likely associated with NGC\,221 (M32), NGC\,205, and Andromeda~IV, based on their positions and radial velocities.  For these objects, we added comments in the catalog. Among them, three H~{\sc ii} region candidates are located in NGC\,205's halo. These may be low-excitation PNe or symbiotic binaries, requiring confirmation via deep spectroscopy.

In M33, the PNe and H~{\sc ii} regions detected by LAMOST are confined to the optical disk. The H~{\sc ii} regions clearly trace the spiral arms, though further verification through chemical analysis and more accurate velocity measurements is needed. The number of H~{\sc ii} regions in M33 exceeds that of PNe, likely due to the larger physical size of H~{\sc ii} regions, which increases the likelihood of detection with LAMOST fibers. In contrast to M31, no PNe or candidates were found in M33's halo, consistent with previous studies \citep{Galera-Rosillo2018}.

\begin{table*}
\begin{center}
\setlength{\tabcolsep}{12.pt}
\caption{\label{tab: cluster} Star Clusters in M31 Identified from the LAMOST Database}
\begin{tabular}{lccccccc}
\hline
\hline
Name &  R.A. & Decl. & RBC $V_{\mathrm{r}}$ & Caldwell $V_{\mathrm{r}}$& $V_{\mathrm{r}}^a$ & APID$^b$ & Status$^c$\\
&  (J2000) & (J2000) & & ($\mathrm{km~s^{-1}}$) & \\
\hline
B001 & 9.96253 & 40.96963 & $-179$ & $-203.3$ & $-209.1 \pm 6.0$ & ~ & 1 \\ 
B002 & 10.01072 & 41.19822 & $-338$ & $-338.2$ & $-336.3 \pm 32.0$ & ~ & 1 \\
B003 & 10.03917 & 41.18478 & $-351$ & $-377.0$ & $-329.5 \pm 11.1$ & ~ & 1 \\
B004 & 10.07468 & 41.37786 & $-369$ & $-369.8$ & $-374.3 \pm 1.4$ & ~ & 1 \\
B005 & 10.08462 & 40.73287 & $-265$ & $-291.6$ & $-292.5 \pm 6.9$ & ~ & 1 \\
... & ... & .. & ... & ... & ... & ... & ... \\
B370 & 11.31000 & 41.96132 & $-352$ & $-352.7$ & $-382.1 \pm 4.7$ & 1773 & 1 \\
B372 & 11.38908 & 42.00678 & $-216$ & $-226.9$ & $-242.9 \pm 5.8$ & 545 & 1 \\
B373 & 11.42436 & 41.75929 & $-219$ & $-216.2$ & $-228.9 \pm 1.4$ & 1920 & 1 \\
... & ... & .. & ... & ... & ... & ... & ... \\
SK168B & 11.10958 & 40.25143 & ~ & ~ & $-67.8 \pm 7.4$ & ~ & 2 \\ 
SK214B & 11.47515 & 39.94646 & ~ & ~ & $-104.8 \pm 9.7$ & ~ & 2 \\ 
SK223B & 11.63737 & 40.11049 & ~ & ~ & $-138.2 \pm 4.0$ & ~ & 2 \\ 
... & ... & .. & ... & ... & ... & ... & ... \\
\hline
\end{tabular}
\end{center}
\tablenotetext{a}{Radial velocity from LAMOST.}
\tablenotetext{b}{Andromeda project identification number in \cite{Johnson2015}.}
\tablenotetext{c}{Classification status (1: confirmed star cluster, 2: candidate star cluster)}
\tablecomments{This table is published in its entirety online only in the machine-readable format.}
\end{table*}

\begin{table*}
\begin{center}
\setlength{\tabcolsep}{3.5pt}
\caption{\label{tab: sg}Supergiants in M31 Identified from the LAMOST Database}
\begin{tabular}{lccccccccccc}
\hline
\hline
Designation &  R.A. & Decl. & $V_{\rm r}$ & $V_{\rm exp}$ & $R_{\rm gal}$ & $P_{\rm M31}$ & $P_{\rm MW}$ & $V$ & $B-V$ & Type$^a$ & W24$^b$ \\
&  (J2000) & (J2000) & (km\,s$^{-1}$) & (km\,s$^{-1}$) & (kpc) & & & (mag) & (mag) & & \\
\hline
J004207.85+405152.1 & 10.53273 & 40.86448 & $-452.8 \pm 6.5$ & $-$411.9 & 10.9 & 0.9012 & 0.0 & 16.988 & 0.761 & ysg & 1 \\ 
J004247.29+414451.0 & 10.69708 & 41.7475 & $-262.6 \pm 3.3$ & $-$225.7 & 18.2 & 0.9185 & 0.0 & 16.410 & 0.465 & ysg & 1 \\ 
J004120.56+403514.6 & 10.33567 & 40.58741 & $-438.0 \pm 4.9$ & $-$439.7 & 15.9 & 0.9999 & 0.0 & 17.099 & 0.599 & ysg & \\ 
J004428.08+415503.0 & 11.11704 & 41.91751 & $-136.2 \pm 1.5$ & $-$117.0 & 13.1 & 0.9768 & 0.0273 & 16.773 & 2.046 & rsg & 1 \\ 
J004731.11+422748.8 & 11.87966 & 42.46357 & $-91.1 \pm 1.5$ & $-$55.6 & 20.4 & 0.9243 & 0.2720 & 17.354 & 2.061 & rsg\_c & \\ 
... & ... & .. & ... & ... & ... & ... & ... & ... & ... & ... & ... \\ 
J003913.02+401655.5 & 9.80425 & 40.28211 & $-504.8 \pm 3.8$ & $-$527.3 & 16.8 & 0.9706 & 0.0 & 18.307 & -0.009 & sg\_c & \\ 
J004235.77+405855.8 & 10.64904 & 40.98217 & $-353.0 \pm 58.3$ & $-$375.0 & 10.0 & 0.9717 & 0.0 & 19.202 & 0.29 & sg\_c & \\ 
J004503.67+413708.5 & 11.26533 & 41.61903 & $-106.9 \pm 6.6$ & $-$142.8 & 10.6 & 0.9258 & 0.1389 & 16.151 & 0.176 & sg\_c & \\ 
... & ... & .. & ... & ... & ... & ... & ... & ... & ... & ... & ... \\ 
J003853.73+403830.2 & 9.72390 & 40.64173 & $-512.8 \pm 6.5$ & $-$474.2 & 17.2 & 0.9113 & 0.0 & 18.441 & 0.143 & star & \\ 
J004205.31+410254.0 & 10.52213 & 41.04835 & $-498.5 \pm 5.7$ & $-$493.2 & 4.1 & 0.9980 & 0.0 & 18.829 & 0.112 & star & 1 \\ 
J004505.51+414658.3 & 11.27296 & 41.78288 & $-67.6 \pm 7.5$ & $-$57.5 & 9.4 & 0.9934 & 0.5734 & 17.096 & 0.276 & star & \\ 
... & ... & .. & ... & ... & ... & ... & ... & ... & ... & ... & ... \\
\hline
\end{tabular}
\end{center}
\tablenotetext{a}{Classification type: known supergiants are labeled with types as reported in literature, and candidates are denoted with ``\_c''. ``sg\_c'' represents new candidates identified in this study. Objects in the ``Stars in the M31 Catalog'' are labeled as ``star''.}
\tablenotetext{b}{Objects labeled with ``1'' indicate common entries identified in \citet{Wu2024}.}
\tablecomments{This table is published in its entirety online only in the machine-readable format.}
\end{table*}

\subsection{Star Clusters} \label{Star Clusters Results}

We compiled a catalog of 344 confirmed star clusters and 36 cluster candidates in M31 after excluding the low-SN spectra as well as those with galaxy-like features.  The known star clusters in M33 were excluded from the catalog. These clusters span a wide range of locations in M31, from the bulge to the outer halo. Table \ref{tab: cluster} lists the names, positions, radial velocities, and classification status of these objects. Most of the sources were drawn from the RBC V5 and the Panchromatic Hubble Andromeda Treasury (PHAT) survey \citep{Dalcanton2012, Johnson2012, Johnson2015}. However, 51 objects classified as GC candidates in the RBC V5, which exhibited large redshifts characteristic of galaxies, were excluded from our final catalog. Another notable case is LAMOST J004251.86+404409.8 (SK055A in RBC V5), previously classified as a confirmed GC, but now reclassified as a background galaxy based on its redshift of $z \approx 0.18$.

Additionally, some spectra from previously known clusters, both young and old, displayed emission lines of varying strength. Those with weak continua and strong Balmer emission lines were reclassified as H~{\sc ii} regions. These regions, often populated by young stars, show slightly extended morphologies and are unlikely to evolve into older, gravitationally bound clusters due to their weaker stellar associations. In the remaining spectra, the weaker emission features are likely due to either the clusters themselves or the diffuse ionized gas present in M31's disk or bulge.

Upon reviewing PAndAS images of a subset of cluster candidates identified through our RFC predictions, we found no new candidates with obvious extended morphologies. Nonetheless, a few compact, unresolved clusters may remain undetected. Future space-based telescopes, such as the China Space Station Telescope \citep[CSST;][]{Zhan2011, Zhan2018}, are expected to shed light on these remaining objects.

\subsection{Supergiant Candidates} 
\label{subsec: supergiant candidates}

Our final catalog of M31 supergiants, identified from the LAMOST spectroscopy, contains 375 objects.  Recently, \citet{Wu2024} carried out a systematic identification of supergiants in M31 and M33 using the LAMOST data, resulting in 199 supergiant candidates in M31 in the ``Rank1" and ``Rank2" classes.  Of these, 86 objects are common in both their catalog and ours.  In comparison with their methods, we adopted the updated position-velocity relation from \citet{Massey2016} and arrived at a more strict criterion in velocity (see Sect.~\ref{subsec:Kinematic_Analysis}).  In addition, we focused our analysis on a smaller region with galactocentric radius $R_{\rm gal}<$25\,kpc.  However, we chose not to apply photometric and astrometric cuts, since the process could cause loss of some targets. 

Based on their findings, we have updated the status of these objects in our catalog, reclassifying 56 common individuals from new candidates to known supergiant candidates. In addition, we incorporated 40 objects from the ``Stars in the M31 Catalog" provided by Nelson Caldwell as known candidates. Due to their high luminosity, which makes them detectable by LAMOST, these objects are likely supergiant candidates. In total, our supergiant catalog includes 183 supergiants and candidates documented in previous studies, as well as 192 newly identified supergiant candidates. We cross-matched our objects with the LGGS photometry catalog \citep{Massey2006} using a radius of 3\arcsec\ to obtain additional magnitude and color information. The basic properties of these objects are summarized in Table~\ref{tab: sg}.

\begin{figure}
\centering
    \includegraphics[width=1.0\columnwidth]{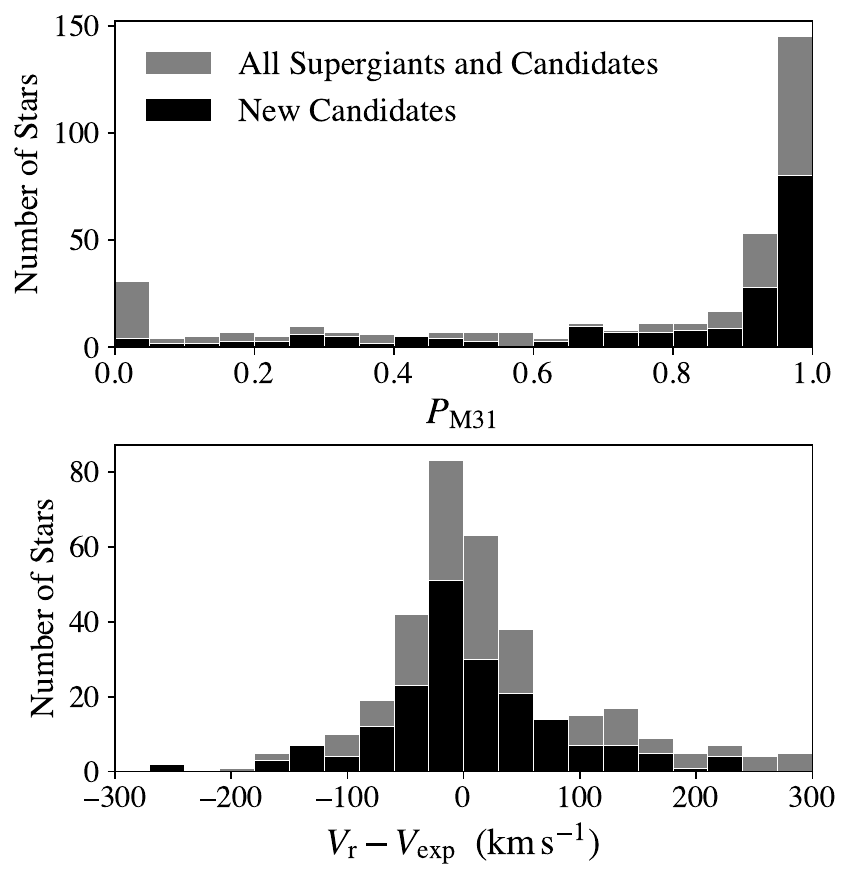}
     \caption{Top: Distribution of the probability of M31 membership ($P_{\rm M31}$, values scaled from 0 to 1) of our newly identified supergiant candidates(in black), along with known supergiants and candidates (in grey).  Bottom: Distribution of ($V_{\rm r}-V_{\rm exp}$) for the known and newly identified supergiants and candidates, with the same color-coding as above.
    \label{fig:sg_probability}}
\end{figure}

Despite that M31 is located along the direction of northern Galactic halo, contamination from the MW foreground stars remains significant.  \citet{Drout2009} restricted yellow supergiant candidates to radial velocities $< -$150\,km\,s$^{-1}$ to estimate contamination of foreground stars.  This approach aligns well with our result of a $3\rm \sigma$ boundary for the MW objects ($V_{\rm r} < -$149.5\,km~s$^{-1}$).  While efficient in filtering out Milky Way disk stars, this criterion is less so for the stars in the Galactic halo.  Since a significant portion of M31's systemic radial velocity is influenced by the relative motion of the Sun, some Galactic halo stars can exhibit large negative radial velocities that mimic M31's motion, complicating our exclusion of these object from the sample.  As such, stars from the Galactic halo are the primary source of contamination in our data. 

To address this, we employed a likelihood-based approach that differentiates between foreground stars and those belonging to M31 by considering both velocity and positional information.  An object is assigned a higher probability of belonging to M31 if its observed radial velocity aligns closely with the expected velocity, which is calculated based on its position in M31's deprojected disk plane. However, the $3\sigma$ boundary is a conservative criterion.  The circular velocities of short-lived supergiants are expected to closely follow M31's rotation velocity, except for a few runaway stars \citep{Evans2015}.  Among the new candidates identified through radial velocity, 152 fall within the $1\sigma$ confidence interval, while 32 lie between the $1\sigma$ and $2\sigma$ interval.  As $P_{\rm M31}$ decreases, the likelihood of contamination from Galactic halo stars increases.  Fig.~\ref{fig:sg_probability} shows the probability distribution of our sample, highlighting that a subset of previously identified objects has a low likelihood of M31 membership.  Caution is advised when interpreting these results.

\begin{figure*}
\centering
    \includegraphics[width=1.8\columnwidth]{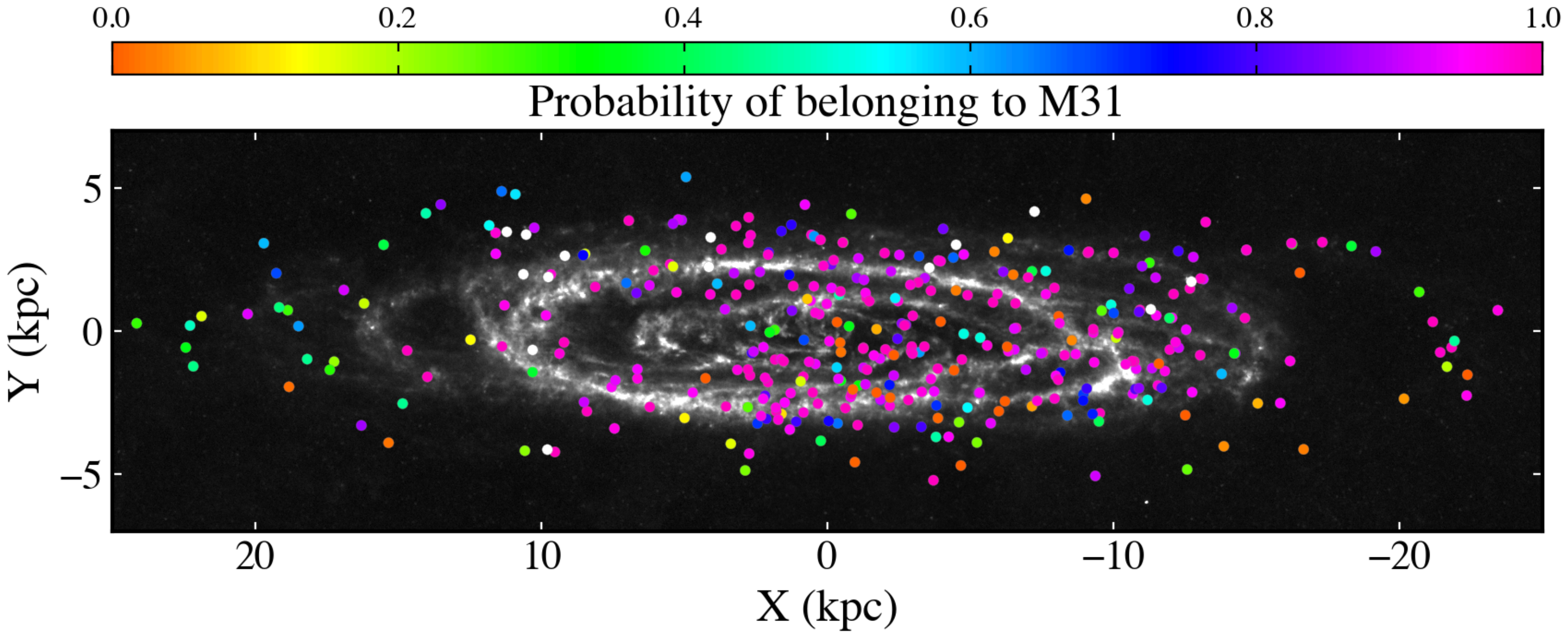}
     \caption{Spatial distribution of supergiants and candidates in M31 as identified from the LAMOST database, overlaid on the \textit{Herschel} SPIRE 250\,$\mu$m image \citep{Fritz2012}.  Objects are color-coded according to the probability of M31 membership ($P_{\rm M31}$, values scaled from 0 to 1); a few objects that lack accurate radial velocity information are shown in white. 
    \label{fig:sg_distribution}}
\end{figure*}

Fig.~\ref{fig:sg_distribution} illustrates the spatial distribution of all known supergiants and candidates, with color indicating their $P_{\rm M31}$ values (ranging from 0 to 1). Seven known supergiants with uncertain velocity measurements, along with objects from the ``Unknown" spectra lacking redshift data, are shown in white. The objects span a galactocentric distance ($R_{\rm gal}$) from 0 to $25\, \rm kpc$ to the center of M31. While most high-probability M31 members are spatially correlated with the spiral arms of M31, we also detect outliers in the outskirts of the disk. Notably, several high-confidence candidates are located in the southwestern quadrant of the disk, extending beyond the main spiral arms. These are likely associated with the arc-like structures seen in infrared and H~{\sc i} images \citep{Fritz2012, Chemin2009, Braun2009}. A few H~{\sc ii} regions are also located in this area (see Fig.~\ref{fig:spatial_distribution}). These objects provide valuable insight into the outskirts of M31, and follow-up high-resolution spectroscopy could help further investigate their nature, shedding light on the chemical evolution and nucleosynthetic history of these extended structures.

Interestingly, the distribution of objects in Fig.~\ref{fig:sg_distribution} shows a higher concentration to the right. This pattern is a consequence of M31's rotation, as objects in the northeastern part of the disk are expected to have radial velocities $V_{\rm exp} > -$150 km\,s$^{-1}$ (i.e., $X/R_{\rm gal} \gtrsim$0.67 in Equation~\ref{expected}). As a result, many of these objects blend with foreground disk stars and were excluded from our final sample to reduce contamination. A promising approach to mitigate this bias involves using spectral features identified through high-resolution spectroscopy, which can more reliably separate M31 supergiants from foreground stars \citep[e.g.][]{Massey2016b, Gordon2016}.

\section{Summary} 
\label{sec: summary}

We employed an RFC model to first classify all LAMOST spectra, categorizing objects into three groups, non-emission stars, star cluster candidates, and emission-line objects.  For the spectra that could not be classified using the RFC model, we conducted detailed visual inspection, focusing on identifying potential emission-line nebulae and supergiant candidates with broad emission lines.  High-quality $g$- and $i$-band images from the PAndAS survey were collected and visually checked for star cluster candidates with RF scores exceeding 0.95.  Subsequently, emission-line nebulae and emission-line stars were distinguished based on their continuum levels in the spectra.  For all emission-line nebulae, we systematically measured radial velocities from prominent nebular emission lines and further classified them as planetary nebulae (PNe) or H~{\sc ii} regions, using emission-line ratio diagnostics.  Using spatial position and the radial velocity as measured from the LAMOST spectra, we compared the observed velocities of the remaining objects with those predicted by M31's rotational pattern.  Two Gaussian functions were constructed to estimate the likelihood of an object being associated with M31 or the MW based on kinematic data.  We then selected new supergiant candidates by applying a set of criteria. 

As a result, we present a comprehensive catalog of 1139 unique objects in M31, observed by LAMOST. These objects are classified into three main categories: emission-line nebulae, star clusters, and supergiants, encompassing both confirmed and candidate members. As part of our survey, we also identified 245 emission-line nebulae in M33. Since M31 is no longer part of LAMOST's current observational plan (Phase 3), this catalog represents a definitive record of all confirmed and candidate members identified from the released LAMOST data on M31. Many of the objects in M31 and M33 were initially identified through photometric surveys (e.g., H~{\sc ii} regions). The addition of LAMOST spectra provides essential complementary data, allowing us to better characterize their nature. The cataloged objects serve as valuable tracers for studying the chemical composition, kinematics, and stellar populations of our neighboring galaxies, offering key insights into their formation and evolutionary history.

\section*{Acknowledgements}
We are grateful to the anonymous referee whose excellent comments and suggestions greatly improved this article. This work is partially supported by the National Natural Science Foundation of China 12173034 and 12322304, the National Natural Science Foundation of Yunnan Province 202301AV070002 and the Xingdian talent support program of Yunnan Province. We acknowledge the science research grants from the China Manned Space Project with NO.\,CMS-CSST-2021-A09, CMS-CSST-2021-A08 and CMS-CSST-2021-B03. 

The emission-line fitting procedure was performed using \texttt{Scipy} \citep{scipy} and \texttt{Specutils} \citep{specutils}.

Guoshoujing Telescope (LAMOST -- Large Sky Area Multi-Object Fiber Spectroscopic Telescope) is a National Major Scientific Project built by the Chinese Academy of Sciences. Funding for the project has been provided by the National Development and Reform Commission. LAMOST is operated and managed by the National Astronomical Observatories, Chinese Academy of Sciences.

\section*{Data availability}
The complete catalogs are published in their entirety online in the machine-readable format, and also available in electronic forms at the CDS.

\bibliography{ms.bib}{}
\bibliographystyle{aasjournal}

\end{document}